\documentclass[%
reprint,
superscriptaddress,
amsmath,amssymb,
aps]{revtex4-2}

\usepackage{graphicx}
\usepackage{dcolumn}
\usepackage{bm}

\usepackage[utf8]{inputenc}
\usepackage{placeins}
\usepackage[normalem]{ulem}
\usepackage{color}

\usepackage{eurosym}

\begin{document}
\preprint{APS/123-QED}

\title{Inferring supply networks from mobile phone data to estimate systemic risk of an economy}

\author{Tobias Reisch}
\thanks{T.R. and G.H. contributed equally.}
\affiliation{Section for Science of Complex Systems, CeMSIIS, Medical University of Vienna, A-1090 Vienna, Austria}
\affiliation{Complexity Science Hub Vienna, A-1080 Vienna, Austria}

\author{Georg Heiler}%
\thanks{T.R. and G.H. contributed equally.}
\affiliation{Complexity Science Hub Vienna, A-1080 Vienna, Austria}
\affiliation{Institute of Information Systems Engineering, TU Wien, A-1040 Vienna, Austria}

\author{Christian Diem}
\affiliation{Complexity Science Hub Vienna, A-1080 Vienna, Austria}
\affiliation{Institute for Finance, Banking and Insurance, Vienna University of Economics and Business, A-1020 Vienna, Austria}

\author{Stefan Thurner}
\email{Corresponding author, e-mail: stefan.thurner@meduniwien.ac.at}
\affiliation{Section for Science of Complex Systems, CeMSIIS, Medical University of Vienna, A-1090 Vienna, Austria}
\affiliation{Complexity Science Hub Vienna, A-1080 Vienna, Austria}
\affiliation{Santa Fe Institute, Santa Fe, NM 85701, USA}

\date{\today}


\begin{abstract}   
National economies rest on networks of millions of customer-supplier relations. Some companies --in the case of their default-- can trigger significant cascades of shock in the supply-chain network and are thus  systemically risky. Up to now, systemic risk of individual companies was practically not quantifiable, due to the unavailability of firm-level transaction data. So far, economic shocks are typically studied in the framework of input-output analysis on the industry-level that can't relate risk to individual firms. Exact firm-level supply networks based on tax or payment data exist only for very few countries. Here we explore to what extent telecommunication data can be used as an inexpensive, easily available, and real-time alternative to reconstruct national supply networks on the firm-level. We find that the conditional probability of correctly identifying a true customer-supplier link ––given a communication link exists–– is about 90\%. This quality level allows us to reliably estimate a systemic risk profile of an entire country that serves as a proxy for the resilience of its economy. In particular, we are able to identify the high systemic risk companies. We find that 65 firms have the potential to trigger large cascades of disruption in production chains that could cause severe damages in the economy. We verify that the topological features of the inter-firm communication network are highly similar to national production networks with exact firm-level interactions.
\end{abstract}

\maketitle

Bilateral interactions between the agents in an economy lead to networks  that dominate practically all aspects of the economy, ranging from
networks of production \cite{fujiwara2010large,dhyne2015belgian},
finance \cite{boss2004network},
distribution \cite{kaluza2010complex}, 
consumption \cite{de2020consumption},
and recycling \cite{schwarz1997implementing}. 
Networks are not only the basis of an efficient functioning of the economy, they are also the source of some of its implied risks and, in particular, systemic risk, or the risk that a large fraction of networks stop to function and do no-longer fulfil their function. Remarkably, the understanding of the economy in terms of its underlying networks has not arrived at mainstream economics \cite{arthur2021foundations}. 

Since about two decades systemic risk has been associated with network structures and ways to quantify it are nowadays available.The main idea behind the quantification of systemic risk is to estimate the economic or financial consequences of a defaulting node or link in a given network on the entire system. The fraction of the total system affected is typically associated with the systemic risk of a node or link. Knowing the systemic risk contributions of agents offers a way to quantify the resilience and robustness of a system.  The first networks available to research were financial networks such as networks of inter-bank claims and liabilities  \cite{boss2004network}, or of  overnight money markets  \cite{iori2008network}. Systemic risk in these networks was first quantified with network measures like betweenness centrality \cite{boss2004contagion}, which were later improved by explicitly incorporating economic default mechanisms and the associated accounting procedures  \cite{battiston2012debtrank, thurner2013debtrank}. Further extensions involved multilayer networks \cite{leon2014financial,poledna2015multi}, overlapping portfolios \cite{pichler2021systemic,cont2019monitoring}, in the context of financial networks, as well as some applications in the real economy \cite{fujiwara2016debtrank}, and lately, also in  production networks \cite{inoue2019firm,diem2021quantifying}.

Systemic risk in mainstream economics has often been discussed not on the basis of networks \cite{adrian2011covar,acharya2017measuring}, but on financial time series data that obviously can't account correctly for cascading processes. It is exactly the cascading that leads to extraordinary large effects that are often associated with the fat tailed distributions of losses \cite{moran2019mays}. The default of Lehman Brothers in 2008 \cite{haldane2009rethinking, longstaff2010subprime}, the 2008-2010 global food crisis \cite{amour2016teleconnected} and, more recently, world wide supply chain disruptions due to the COVID-19 pandemic \cite{rowan2020challenges,mahajan2021covid} are only a few examples of severe events in financial markets, basic provision, or production networks, where cascading plays an essential role. 

A network-based quantification of systemic risk makes it possible to identify the weak points in these systems and consecutively allows one to design mitigation strategies, for example an adaptive systemic risk tax to reduce the systemic risk in a banking system \cite{poledna2016elimination,leduc2017incentivizing} or the computation of optimal networks that carry a minimum of systemic risk \cite{diem2020minimal, pichler2021systemic}.
However, the  computation of systemic risk requires the detailed understanding of the structure and dynamics of the underlying networks, which hitherto posed a major challenge \cite{brintrup2018supply}.

\begin{figure*}[t]
\centering
\includegraphics[width=0.99\textwidth]{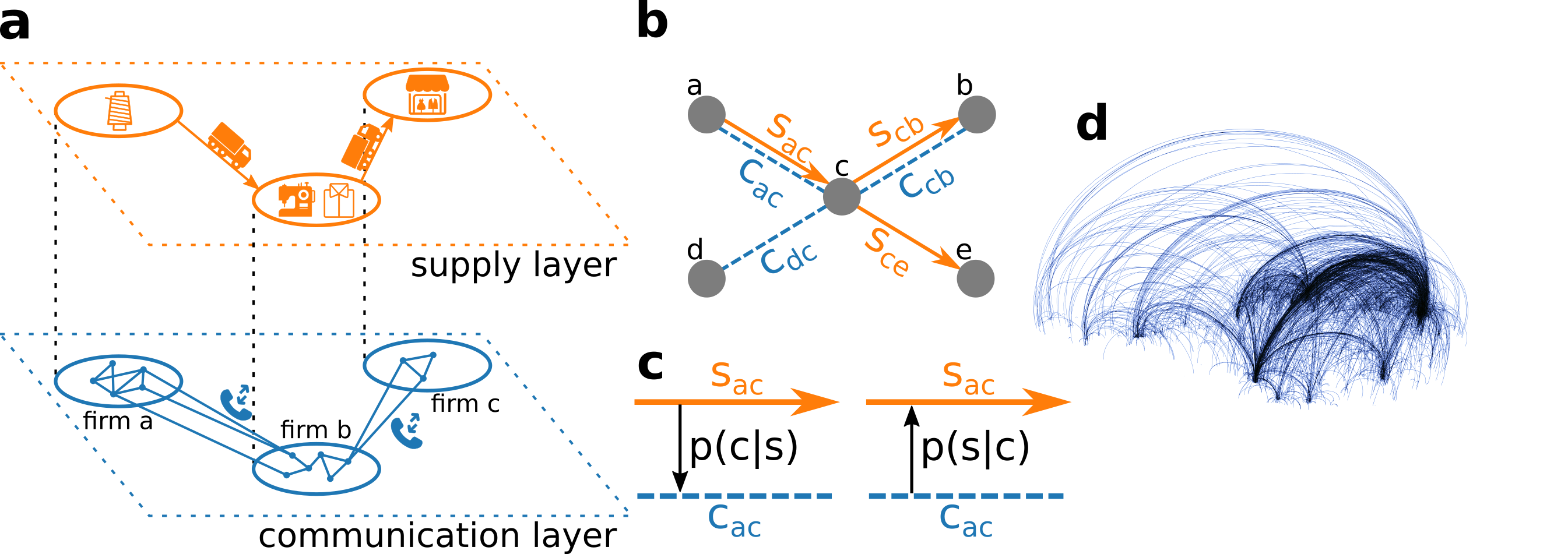}
\caption{(a) Schematic view of the inter-firm multilayer network with a communication layer (blue) of phone calls between groups of devices that are associated to firms and the supply layer that captures the actual flow of goods (orange). (b) Section of the multilayer network where communication links, $c_{ij}$, exist if at least one phone call between firms $i$ and $j$ takes place and supply links, $s_{ij}$, exist if goods flow from $i$ to $j$. (c) Conditional probabilities between supply links and communication links are defined as the probabilities to find a supply link, conditional on a communication link being present, $p(s|c)$. (d) The inter-firm communication network as provided by a mobile phone company. Arcs link firms that have an average call duration of more than 150s/d. Firms are slightly dislocated randomly, enough to ensure the anonymity of  companies. }
\label{fig:call_nw_explanation}
\end{figure*}  

This is particularly true for systemic risk in production networks. Only for very few countries buyer-supplier relations are known on a granular level of individual companies from which the supply-chain networks can be constructed. For Hungary value-added tax (VAT) data exists that specifies which company pays VAT to another. From this the {\em exact} national supply-chain network has been reconstructed \cite{borsos2020unfolding}, containing more than 89,000 companies and 235,000 business relations (links). Using estimations for production functions for these companies makes it possible to obtain the national production network. Using this as an input, firm level systemic risk for all individual companies were computed by using an appropriately designed SR measure, the  Economic Systemic Risk Index (ESRI) \cite{diem2021quantifying}. It is a network-based measure to estimate the fraction of the total production output (goods and services) of the economy that is affected by a firm's (short-term) failure. 

However, the Hungarian data is an exception. Granular and exhaustive datasets on the supply network of an entire nation are notoriously hard to obtain. Data exists only for a handful of countries, Japan \cite{fujiwara2010large}, Belgium \cite{dhyne2015belgian}, Brasil \cite{silva2020modeling}, and Hungary \cite{borsos2020unfolding}. Customer-supplier relations are inferred either from surveys and business intelligence \cite{fujiwara2010large}, payment system data \cite{silva2020modeling}, or VAT data \cite{dhyne2015belgian, borsos2020unfolding}. Survey data is typically very costly to collect and suffers from being outdated, highly incomplete, unweighted, and hard to verify \cite{brintrup2018supply}; on the other hand, payment system and tax data --in countries where it is collected-- is  sensitive and access is highly restrictive.
    
In this work we propose an alternative approach to reconstruct the supply-chain network by using the multilayer network structure of firm-to-firm relations. We assume that companies that communicate with each other also entertain customer-supplier relations. We thus focus on two network layers, the flow of goods and services that constitute supply relations and the mobile phone communication between companies. Figure \ref{fig:call_nw_explanation}a schematically depicts the two-layer network. The communication layer (blue) shows the mobile devices belonging to one firm, calling devices in other firms. The supply layer (orange) represents the flow of intermediate products (or services) between firms. In Fig. \ref{fig:call_nw_explanation}b we show the same situation by showing a communication link $c_{ij}$ (blue) between firm $i$ and $j$ if they had at least one phone call within a certain time period and a supply link $s_{ij}$ (orange) if goods or services flow from firm $i$ to $j$. Note that communication links are undirected, supply links are directed.    
    
The coordination of a customer-supplier relation, such as ordering, negotiating prices, or organizing shipping, requires communication between firms and has been studied intensively \cite{humphreys2004impact,paulraj2008inter}. We thus expect the existence of strong link-correlations between the communication and supply layers. From the multilayer network in Fig. \ref{fig:call_nw_explanation}b we define the conditional probability,  $p(s_{ij}|c_{ij})$, to find a supply link, $s_{ij}$, between firms $i$ and $j$ given that a communication link, $c_{ij}$, exists, and vice versa, the conditional probability, $p(c_{ij}|s_{ij})$, to observe a communication link given that a supply link exists, see Fig. \ref{fig:call_nw_explanation}c.
    
Albeit strong legal regulation telecommunication data has been accessible to researchers since more than a decade. Mobile phone data in the form of call detail records (CDRs) that are collected by mobile phone operators for billing purposes have been used to study communication networks and the behavior of millions of people \cite{blondel2015survey}, leading to spectacular insights into the structure of human communication and organization \cite{onnela2007structure,eagle2010network}, human behavior in emergency situations \cite{bagrow2011collective}, the spread of infectious diseases \cite{brdar2016unveiling, jia2020population} and the principles of human mobility \cite{gonzalez2008understanding,song2010limits,schlapfer2021universal}. CDRs allow for population-wide coverage, granular resolution of interactions on the person level, and the possibility to be combined with information, such as age and gender. Even though possible, inter-firm or organisational networks have so far not been studied systematically with mobile phone data.

\begin{figure}[ht]
\centering
\includegraphics[width=\linewidth]{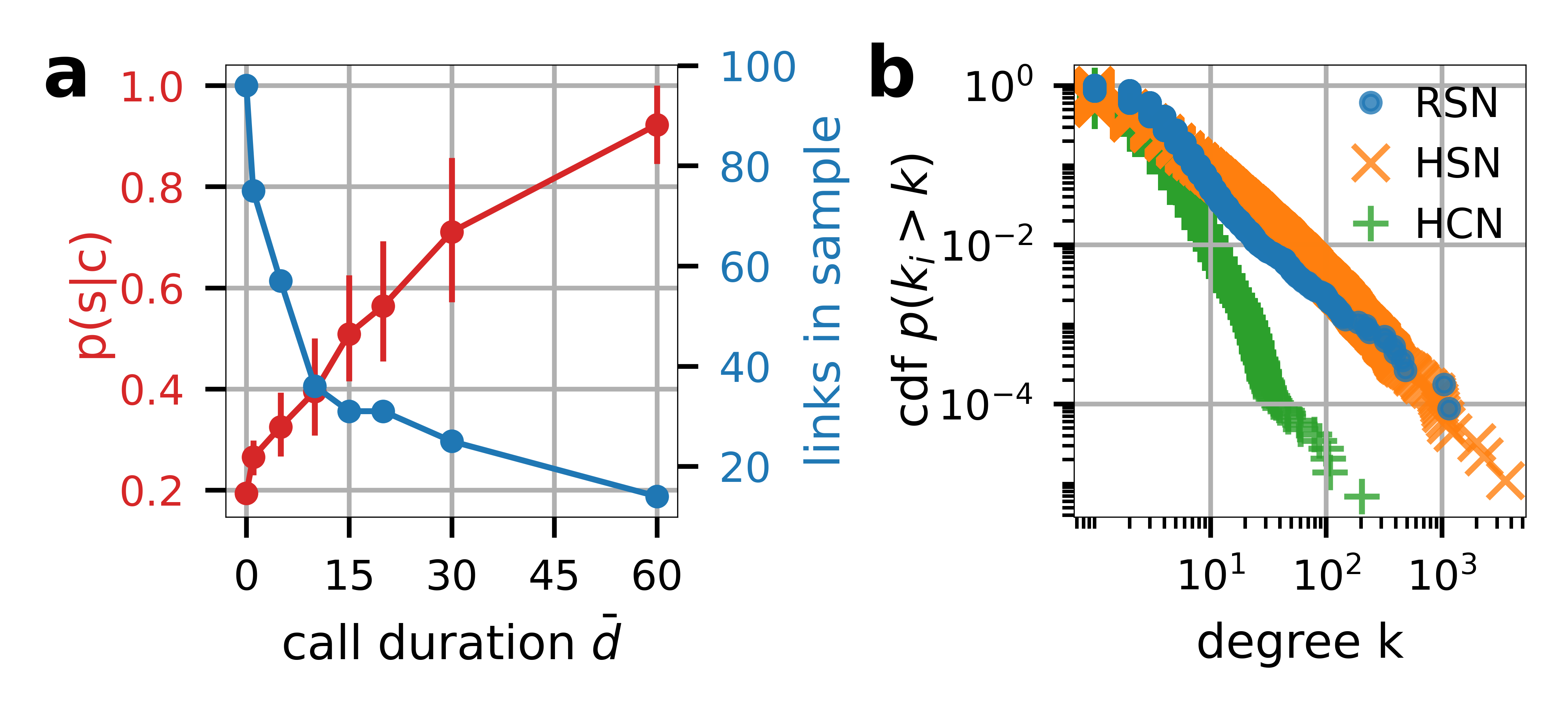}
\caption{(a) Probability $p(s|c)$ to find a supply link, $s_{ij}$, given that there exists a communication link, $c_{ij}$, between firms $i$ and $j$ for communication links exceeding a given call duration, $\bar{d}_{ij}$. Error bars denote the quartiles of a bootstrap simulation described in SI Text 1.
(b) Cumulative distribution function $p(k_i>k)$ for the degree $k$ of the RSN (blue dots), HSN (orange x's) and HCN (green pluses). The degree distribution of the HSN is much more similar to the RSN than the HCN. Errorbars denote the quartiles of a bootstrap simulation described in SI Text 1.}
\label{fig:layer_similarity}
\end{figure}

Through a cooperation with a large mobile phone provider we have access to a dataset of CDRs in a medium-sized European country that allows us to identify groups of phones that are associated with a company through anonymized billing information, for details, see Materials and Methods. The dataset contains additional information on the firm's primary industry classification and balance sheet information. In Fig. \ref{fig:call_nw_explanation}d we show the corresponding firm-to-firm communication network (FCN) as obtained from our data. Firm locations are shifted by random distances (on average 30km) to ensure the anonymity of companies. 
Arcs in the figure represent communication links between firms. We find many short-range interactions within one city or economic region and few long-range interactions. We are intentionally vague with regards to information concerning the mobile phone provider because we are contractually bound to ensure its anonymity, as well as to protect sensitive business information such as the exact market share in the business-to-business market.

Here we demonstrate that phone data can indeed be used to reasonably reconstruct supply networks that allow for a meaningful estimation of firm-level economic systemic risk of an economy. The method is an efficient  alternative to survey, tax, or bank transactions estimates. It uniquely allows us to study supply networks and monitor economic systemic risk in real time and provides a nearly complete overview of a nation's production network.

\section*{Results}

{\bf Conditional supply-link probability.}
We determine the conditional supply-link probability $p(s|c)$ by comparing the firm communication network, shown in Fig.   \ref{fig:call_nw_explanation}d, with ground-truth information on the real  customer-supplier relations, obtained from a nation-wide survey in April 2020. In the online survey more than 100,000 companies and businesses were asked to share their ten most critical suppliers and customers, respectively. More than 5,900 firms declared at least one supplier or customer with a total of more than 17,000 customer-supplier relations reported. For details on the survey, see SI Text 1. We obtain the overall probability that a supply link exists between two companies, given that they had at least one conversation event in the observed time period of approximately 150 days, is $p(s|c) = 0.19$. For the conditional communication probability we get $p(c|s) = 0.27$. For comparison, the respective marginal probability from the firm communication network directly is $p(c)=0.002$. For the linking probability ––using  Hungarian data––  we get $p(s)=0.00005$. Since both values are orders of magnitude smaller than the conditional probabilities, highly significant link correlations between the supply and communication layers are indicated.

The conditional link probability increases with the intensity of the firm-firm communication. As a proxy for the latter we use the average daily call duration, $\bar{d}_{ij}$, in seconds per day. In Fig. \ref{fig:layer_similarity}a $p(s|c)$ is shown as a function of $\bar{d}_{ij}$ (red). The number of links used to calculate the overlap is shown in blue. $p(s|c)$ rises from 19\% to values around 70\% for $\bar{d}_{ij}=30$s/d and around 90\% for 60s/d. The number of links reduces from 75 to 14 links as $\bar{d}_{ij}$ increases. Note that errors do not increase, because a higher probability is associated with a smaller error. For details of the computation and errorbars, see SI Text 1.

For the supply network ($p(c|s)$) the best proxy for tie strength would be the amount of traded goods. However, this information is not available, so we estimate the link weight as the product of the firm's sizes. Here, to stay consistent on the communication data, we proxy the firm size with the number of devices associated with a firm. Supplementary figure \ref{fig:SI_psc} shows $p(c|s)$ for the networks thresholded by the number of devices per firm in red and the number of links in the underlying sample in blue. We find an increase from 27\% to around 60\% for the network of firms with 4 or more devices. For thresholds larger than 4, the curve levels off and stabilizes around 70\% for thresholds of 6 or more devices. The number of links drops as in Fig. \ref{fig:layer_similarity}a, but again, the error-bars are still sufficiently small.
\\

{\bf Reconstructing the supply network.}
For obtaining an estimate of the supply network, based on the FCN, we chose $\bar{d}_{ij}=30$s/d, with the aim to balance the loss of information due to ignored supply links and increasing link correlations due to the thresholds. This particular threshold is the result of a minimization of the Kullback-Leibler divergence for degree distributions of the HSN and thresholded FCNs, described in SI Text 2. We arrive at an unweighted and undirected reconstructed supply network (RSN). To get an estimate for the link directions (firm $i$ supplies $j$ or vice versa), we use classical input-output tables of the national statistical office. They contain information on the volume of trade between economic sectors in the economy. An element of the input-output table,  $G_{ab}$, describes the flow of goods (in Euro) from sector $a$ to sector $b$. We denote the number of links (firm-firm supply relations) from sector $a$ to sector $b$ by $L_{ab}$ and assume that the ratio of links from one sector to the other is proportional to the ratio of goods flowing between these sectors, ${L_{ab}}/{L_{ba}} \approx {G_{ab}}/{G_{ba}}$. For example, the flow between the agricultural sector ($a$) and the food industry ($f$) is $G_{af} \approx 3,400 \mathrm{ m \text{\euro}}$, while the food industry sold goods for $G_{fa} \approx 450 \mathrm{ m \text{\euro}}$ to the agricultural sector. We now assume that it is ${3,400}/{450} \approx 7.6$ times more likely that a supply link points from a firm $a$ to one in $f$. We now consider every link from firm $i$ in sector $a$ to firm $j$ in sector $b$ in the RSN and assign it a direction according to the probability 
\begin{equation}
    p(i \rightarrow j) = \frac{G_{ab}}{G_{ab}+G_{ba}} \, .
    \label{eq:pij_iot}
\end{equation}
Since we perform this assignment stochastically, we should think in ensembles of RSNs. Finally, we estimate a supply-link weight for every link in the RSN. We use the companies' total assets, calculated from the balance sheets, as size information, $s_i$; it is obtained from a commercially available business intelligence database, see Materials and Methods. Following the philosophy of  ``gravity models'' in economics, we assume that large and small firms typically trade large and small volumes, respectively \cite{anderson2011gravity}. Therefore we obtain a link weight estimate between firms $i$ and $j$ as the product of firm sizes, $W_{ij} = s_i s_j$. We will use only relative weights in the following. 
\\

\begin{figure*}[ht]
	\centering
	\includegraphics[width=0.99\textwidth]{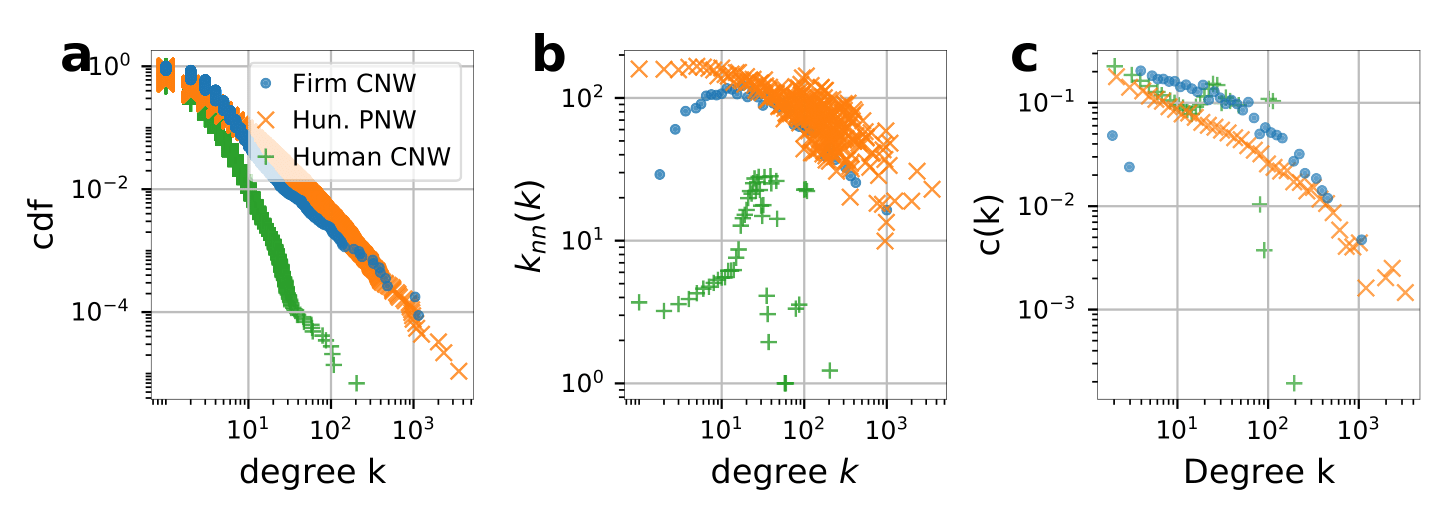}
	\caption{Similarity of the communication and supply layer of the inter-firm multilayer network. (a) Counter cumulative distribution function $p(k>x)$ for the degree $k$ of the FCN (blue dots), HSN (orange x's) and HCN (green pluses). (b) Average nearest neighbor degree $k_{nn}(k)$ and (c) local clustering coefficient c(k) for the three networks. Especially the large-k behavior of the FCN and the HSN is very similar compared to the HCN.}
	\label{fig:SI_layer_similarity_all}
\end{figure*}

{\bf Comparing network topologies of supply-chains, firm-firm communication, and human communication.} It is enlightening to compare the network topology of the so-obtained RSN (blue) with the topologies of the Hungarian supply network (HSN) (orange) (for which exact topology is known \cite{diem2021quantifying}) and the private communication network between individual people (green) (i.e. not between companies). Figure \ref{fig:layer_similarity}b shows the degree distribution of the RSN (blue) in comparison to the exact Hungarian supply network (HSN) derived  from VAT data \cite{borsos2020unfolding}. Both networks are similar and fat tailed, in contrast to the human communication network (HCN) that was obtained from the mobile phone  data set. The RSN has an average degree of $\langle k^{RSN} \rangle = 4.79$. Its degree distribution has a maximum at $k^{RSN}=2$ and its fat tail can asymptotically be approximated by a power law exponent $\alpha_k^{RSN} = 2.18(12)$ for $k^{RSN}>30$. The HSN does not show an increase for small $k$ but also exhibits a fat tail with $\alpha_k^{HSN} = 2.40(3)$, for $k^{HSN}>30$. The average degree is  $\langle k^{HSN} \rangle = 2.1$. For the HCN we find an average degree of $\langle k^{HCN} \rangle = 4.75$. There the decrease of $p(k)$ for high values is stronger, with an exponent of $\alpha_k^{HCN} = 4.89(26)$ for $k^{HCN}>20$. For a deeper comparison of network characteristics, such as clustering coefficient, and nearest neighbor degree, see SI Text 3 and SI Fig. \ref{fig:SI_layer_similarity_all}. 
\\

\begin{figure}[t]
\centering
\includegraphics[width=0.45\textwidth]{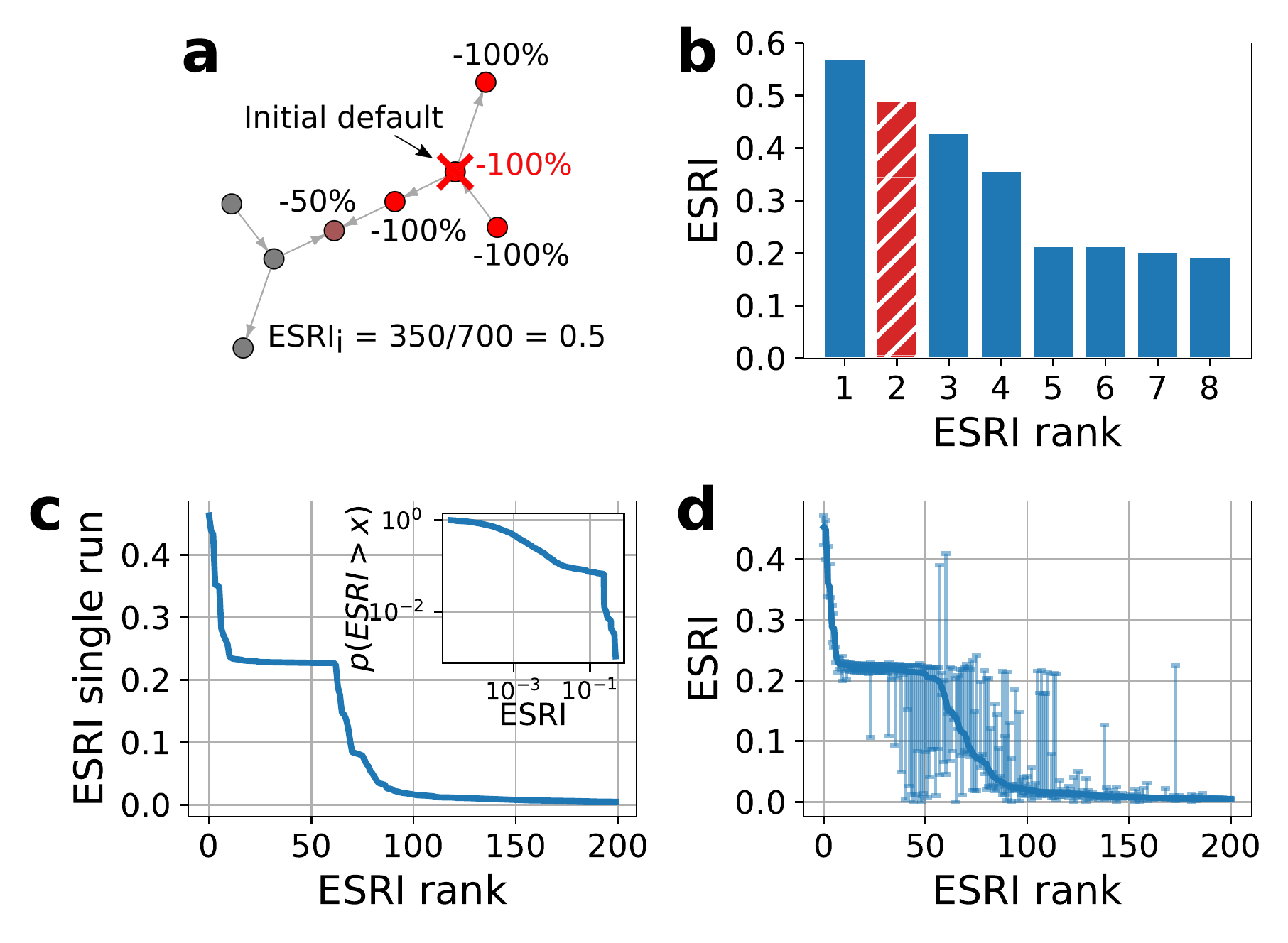}
    	\caption{Economic systemic risk in production networks. 
    	(a) Toy example of a production network with eight firms of the same sector and size. After the default of an initial node (red X) its customers (suppliers) have to reduce their production level according to the share of inputs (supply) they lost. This logic is iterated until a stable configuration is reached, the relative share of economic activity (production) lost is the initial node's ESRI. 
    	(b) The systemic risk profile for the toy network in panel (a), the initial node in panel (a) is highlighted in red. The bars show the ESRI for all firms sorted according to ESRI, from highest to lowest. 
    	(c) Rank ordered systemic risk profile (most risky to the left) for one realization of the RSN. There are 65 high systemic risk firms forming the visible plateau, and a rapid decrease in ESRI for higher rank firms. The maximum $ESRI$ is $0.47$, the majority of the plateau has an $ESRI$ of around $0.21$. The inset shows the counter cumulative distribution function $p(ESRI>x)$ in double logarithmic scale, cropped for $ESRI<10^{-5}$. 
    	(d) ESRI profile for 100 realizations of the RSN. For every firm we show its median ESRI of 100 RSNs, the firms are ranked according to the median ESRI (solid blue line), the error bars show the 25\% and 75\% percentile. For the median ESRI we find a slightly smaller core of around 50 high systemic risk firms with the majority of the risky firms having a median ESRI of around 0.21 and a maximum median ESRI of 0.45. The error bars are small for high and low systemic risk firms, but large for firms in-between, suggesting that their ESRI strongly depends on the direction of one or a few links.}
\label{fig:ESRIspreadMain}
\end{figure}

{\bf Economic Systemic Risk.}
With a reasonable reconstruction of the supply network, RSN, we turn to the quantification of economic systemic risk in the national production network. For quantification we use the economic systemic risk index (ESRI) as developed in \cite{diem2021quantifying}. The algorithm of the ESRI is sketched in Fig. \ref{fig:ESRIspreadMain}a in an example with seven firms of equal size within the same industrial sector. The ESRI of firm $i$, assumes that if firm $i$ (red cross) cannot operate for some time (e.g. defaults) it neither supplies nor demands inputs. The customers and suppliers of $i$ then  reduce their production accordingly, causing a successive reduction of production in their customers and suppliers. This recursive reduction converges to a state where all firms have reduced their level of production in response to $i$'s default. Figure \ref{fig:ESRIspreadMain}a shows the relative reduction for every firm. The fraction of total economic activity lost is the $ESRI_i$ of firm $i$. We use the ESRI algorithm with a generalized Leontief production function that captures the different nature of the production process for companies producing physical goods and companies providing services. Firms with a NACE classification up to F43 are considered to having a physical production process and, hence, are more susceptible to production stops following from  input shortages. For a detailed explanation of the use of generalized Leontief production functions in the ESRI definition, we refer to SI Text 4 and \cite{diem2021quantifying}.

We compute ESRI for every firm in the network and plot their values according to their rank, from highest ESRI to lowest, in Fig. \ref{fig:ESRIspreadMain}b. This is called the {\em systemic risk profile} of the production network. The ESRI for the defaulting firm in panel a is highlighted as the red bar. Performing the same steps for all firms in one realization of the RSN yields the systemic risk profile shown in Fig. \ref{fig:ESRIspreadMain}c, where we  show the 200 riskiest firms. The profile shows similar characteristics to what has been reported for the exact production network of Hungary \cite{diem2021quantifying}, namely, a plateau containing the 65 most risky firms, which all, except for a few extremely risky firms, have a similar risk of around $\mathrm{ESRI} \approx 0.21$, followed by a sharp decline for firms that are not part of the plateau. The inset in Fig. \ref{fig:ESRIspreadMain}c shows the cumulative distribution (CDF) $p(\mathrm{ESRI} > x)$ of the ESRI in log-log scale. 
    
To take the stochastic nature of the RSN into account we repeat the ESRI calculation. We consider five realizations of the RSN to calculate their mean ESRI. Subsequently, due to computational challenges, we focus on the $1000$ most risky firms only, after ranking them according to their mean ESRI. For those we repeat the ESRI calculation $100$ times. For each node we get a distribution of ESRI values. Figure \ref{fig:ESRIspreadMain}d shows the median ESRI for every firm as a solid line; the 25\% and 75\% quantiles are indicated by the errorbars. An alternative way to investigate the ESRI profile of the RSN is to plot the maximal systemic risk of every node. This method yields similar results and is shown in SI Fig. \ref{fig:SI_ESRImax} in SI Text 5. The median ESRI per node profile in Fig. \ref{fig:ESRIspreadMain}d shows the same characteristics as the single run in Fig. \ref{fig:ESRIspreadMain}c, a plateau of high-risk firms and a rapid decline of ESRI outside of the plateau. In contrast to the single run ESRI profile, the plateau consists of only around 50 firms. The spread of the ESRI distributions for individual nodes is small for high- as well as low-risk nodes, indicating that the results are remarkably stable and robust. For the intermediate risk firms  error-bars become large, indicating that their ESRI depends on the direction of one or few links. It is a well-known feature of systemic risk and the ESRI that single links or link directions can have a large influence \cite{thurner2013debtrank, diem2021quantifying}. Ref. \cite{diem2021quantifying} explains that some nodes ``inherit'' systemic risk by being a crucial supplier to a firm that is inherently risky due to e.g. its size. Therefore flipping a link-direction can turn a node from a crucial supplier of a central firm to a buyer of that firm, which strongly reduces its inherited systemic risk. 

\begin{figure}[t]
\centering
\includegraphics[width=0.75\linewidth]{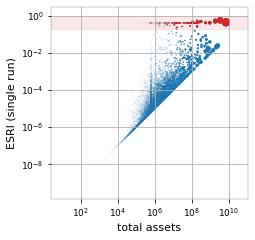}
\caption{ Economic systemic risk vs. firm size measured as total assets in log-log scale. Marker size represents total assets; companies in the ``plateau'' of Fig. \ref{fig:ESRIspreadMain}c are marked red and highlighted by red shading. For a given firm-size there is an obvious lower bound for the ESRI that corresponds to the firm-size (here no firm can loose less than all assets when it defaults). Although the correlation between size and ESRI is high, we find small and large firms in the ``plateau", suggesting that firm-size is not a good tool to identify high-systemic risk firms.}
\label{fig:ESRIvsSize}
\end{figure}

To understand which firms are in the plateau of Fig.  \ref{fig:ESRIspreadMain}c, in Fig. \ref{fig:ESRIvsSize}, we plot the firm-size, approximated by the firms' total assets against the ESRI. It is evident that the high systemic risk plateau (highlighted in red) contains large and small firms, with their total assets spanning more than 4 orders of magnitude. Although firm size correlates well with ESRI (Spearman's $\rho=0.87$), it is not a good predictor for systemic risk since for a given firm-size the ESRI can vary by several orders of magnitude. A similar situation is described in \cite{diem2021quantifying}.
 
The 65 firms found in the high systemic risk plateau mainly belong to the manufacturing sector (NACE lvl. 1 category C, 77\%), followed by companies in the electricity, gas stream and air conditioning supply (D, 8\%) and financial and insurance activities (K, 6\%) sectors. The full composition is listed in SI Tab. \ref{tab:sector_composition_lvl2}. In contrast to the exact Hungarian production network \cite{diem2021quantifying}, several companies from non-manufacturing sectors (NACE $\ge$ 45) are found in the plateau. This is somewhat unexpected since they are associated with linear production functions (see SI Text 4), which causes their shock spreading behavior to be less extreme than for Leontief producers.
    
{\bf Robustness of results.}
Our study is subject to several limitations, in particular 
    (i) the imperfect overlap of the two communication and supply-link layers, limiting the possible accuracy, 
    (ii) the limited market coverage of the phone provider (resulting in limited agreement even if $p(s|c) = 1$), see SI Text 6, and 
    (iii) errors originating from the network reconstruction uncertainties in the estimations of directions and weights. 

To estimate the biases and errors introduced by these weaknesses, we perform several simulation studies.
First, we generate a synthetic communication network based on the HSN and the probabilities to find a communication link, where a supply-link is present $p(c|s)$, and where no supply-link is found $p(c|\lnot s)$. 
From this synthetic communication network we then take a sample of nodes according to an estimated market share $m$ of the data provider and calculate the induced subgraph comprised by links only between the sampled nodes. 

Finally, following the procedure used on the empirical data, we reconstruct a supply network from this synthetic communication network and calculate the ESRI. We calculate Spearman's rank correlation coefficient, $\rho$, between the ESRI as calculated on the full, real HSN and on the reconstructed subgraph. After repeating these steps for $100$ times with $m=1/3$, $p(c|s) = 0.21$ and $p(c|\lnot s) = 9.3 \times 10^{-5}$,  we find an average Spearman correlation of $\langle \rho(ESRI_{HSN}, ESRI_{reconstr})\rangle = 0.563(6)$. In SI Text 6 we address the shortcomings mentioned above one by one and discuss the expected magnitude of the introduced errors. We find that the most relevant effect is caused by the limited market share with a drop of correlation of $\Delta \langle \rho \rangle = 0.31$, followed by the limited overlap, adding another, $\Delta \langle \rho \rangle = 0.13$. The effects from network reconstruction reduce the correlation by only $\Delta \langle \rho \rangle = 0.0004$, which is remarkably small. We calculate the probability that a node that is among the 0.1\% riskiest nodes of the subsample is also among the riskiest 0.1\% of {\em all} nodes and find 32.9(82)\%. The probability that one of the top 0.1\% of the subsample nodes is among the top 1\% of the full network is 47.7(99)\%.

\section*{Discussion}
We show that mobile phone metadata can be used to reasonably reconstruct the flow of goods between firms in an economy, i.e., the supply network. The reconstruction is possible because of the similarity of the communication- and the supply layer of the inter-firm network. This method is one of the very few alternatives to obtain a comprehensive view on national supply network, when there is no VAT or payment system data.

Based on the supply network we calculate economic systemic risk and find that a small core of about 65 high systemic risk firms have the potential to affect large parts of the economic activity. Apart from these core firms systemic risk of companies is generally small. These results agree well with the previous results for Hungary, where a core of 32 high systemic risk firms was found to contribute to 45\% of the overall systemic risk \cite{diem2021quantifying}. With a series of robustness checks we demonstrate the reliability of the results.

Using a large-scale survey on the actual customer-supplier relationships between companies, we find the probability of a supply link to exist, given an existing communication link as $p(s_{ij}|c_{ij}) \approx 0.19$. When thresholding for higher interaction strength of the communication  relation $p(s_{ij}|c_{ij})$ the conditional probability increases strongly to 92\%. Note that the survey asked for the firms' {\em most} critical suppliers. It is almost certain that in the FCN we observe connections to suppliers that are perhaps important but were not classified as critical in the survey, causing $p(s_{ij}|c_{ij})$ to be underestimated. Landline phones are still common practice in many firms; these communication links are not covered, thus further underestimating the overlap of communication and supply links.

We find that the degree exponents of the reconstructed supply network, $\alpha_k^{RSN} \approx 2.18$, and the exact Hungarian supply network, $\alpha_k^{HSN} \approx 2.40$, are similar; the degree exponent of the human-human communication network is much larger, $\alpha_k^{CN} \approx 4.89$. Also for the average nearest neighbor degree and the local clustering coefficient the topology of the RSN is more similar to the topology of the exact HSN than to the HCN.

We showed that the FCN and the HSN are most similar when thresholding communication strength to $d_{ij}>30\mathrm{s/d}$. We sample supplier directions using external information on companies' industry sectors and from input-output tables. Link weights are estimated by the product of firm sizes. Future improvement of the reconstruction method could be reached by using additional information contained in the FCN, such as asymmetries in the calling behavior, temporal patterns in the sequence of calls, as well as using dependencies of supply link weights on communication intensity.
    
Because the reconstruction process is stochastic, we calculate an ensemble of systemic risk profiles and investigate the inter-quartile ranges. For high- and low-systemic risk firms the inter-quartile ranges are small, indicating that results are stable. However, for firms of intermediate systemic risk the inter-quartile ranges are relatively large, indicating that risk changes strongly with the direction of one or a few links. This agrees well with previous results for the Hungarian supply network, where approximately a third of the riskiest firms were found to constitute the periphery of the high systemic risk core. These firms `inherit' the high systemic risk status of important firms by being critical suppliers to these firms \cite{diem2021quantifying}. Also in banking networks it is well known that individual links may dramatically increase systemic risk \cite{thurner2013debtrank,diem2020minimal}. Although the ESRI correlates strongly with firm size, the high systemic risk core is not predicted well by firm size.
    
The method has several limitations. We systematically investigate the error introduced by the imperfect overlap of the communication- and supply layers, the limited market share of the mobile phone provider, and the reconstruction of the link directions. In a simulation study we find an average rank correlation between the true ESRI in the HSN and the ESRI on a carefully simulated synthetic firm communication network of $\langle \rho(ESRI_{HSN}, ESRI_{reconstr}) \rangle = 0.563$. The limited market coverage and the imperfect link overlaps contribute most of the effect. We expect $\langle \rho(ESRI_{HSN}, ESRI_{reconstr}) \rangle$ to be higher in reality since it is based on the estimate for $p(s|c)$ that is a lower bound. Further, despite the limited correlation, our method allows us to capture heterogeneity in shock spreading well and uncovers the localized effects of up- and downstream cascades on the firm level that traditional methods such as input-output models cannot describe.
 
There are also three limitations that could not be addressed explicitly. First, firms use many more communication channels than mobile phones such as landlines, e-mail or physical mail, and a growing number of new interaction channels, such as social media or online portals. Nevertheless, we assume that, if the supply relation is sufficiently strong, firms become more and more likely to use mobile phones to arrange spontaneous meetings, inform partners about delays, coordinate the quality, quantity and timing of deliveries, fix dates, provide support, etc. 

Second, due to the anonymity of the telecommunication data it is not possible to perform targeted surveys on the customers of the phone provider. To reach significant overlap of the survey respondents and the customers of the phone provider, untargeted surveys need a response rate of a considerable fraction of firms within a country.

Third, another consequence of the anonymity of the data is that --by definition-- firms cannot be identified and concrete policy statements can only be made on the level of the network. However, within the anonymity constraints, the effect of heterogeneous shocks in relation to economic sectors and geography can still be investigated. This is important since recent work has shown that heterogeneity in the initial economic shocks can cause dramatically different economic outcomes \cite{inoue2019firm, diem2021heterogeneity}.
    
Since mobile phone data is easily available, the presented method to reconstruct a national production network is cheap, scalable, and easily implemented. The method also captures international links which allow us to identify economic exposures to specific countries. Maybe one of the most interesting features of the method is its temporal resolution, supply relations can be monitored in real-time. This offers the possibility to study how firm-ties form and rewire on the network-level. Observing the restructuring processes of the economy during natural disasters or economic crises take place might become relevant for acute crisis management.
    
\section*{Materials and Methods}
{\bf Data.}
The anonymized (but fine-grained, device-level) call detail record (CDR) data is mapped to an anonymized ID for each company. The observation period is approximately 125 days in autumn 2020 between two lockdowns. The obtained edge list is aggregated for the whole observation period, grouped by each source/destination, anonymized firm ID tuple and the call duration (in seconds) for each arc is summed up. Further, node-level statistics i.e. the number of devices is aggregated. We also calculate a rough location as the centroid of the night-locations of the individual devices. The night-location was previously calculated as described in \cite{covidBasicCSH} for each device.
	
The firm communication dataset is merged with a commercially available business intelligence database that includes balance sheet information the industry classification in the NACE 2008 system  \cite{naceclassification2006}). For details on the anonymization procedure see SI Text 7.
For analysis, we drop NACE J61, J62, M70, and N82 to exclude businesses such as call-centers that have telephone activity at the center of their business and would confound the study with exceptionally high numbers of calls.
	
To compare the reconstructed supply network (RSN) with a real supply network we use a dataset based on granular VAT reporting in Hungary (HSN), described in \cite{borsos2020unfolding,diem2021quantifying}. It contains a link between two firms only if at least two transactions occur in two different quarters. We use the data from 2017, where only transactions with a tax content larger than 1,000,000 Forint (approx. 3000€) are included. Hungarian VAT rates range from a 27\% base rate to a 18\% and 5\% reduced tax rate for certain foods, pharmaceuticals, etc. and a 0\% rate for public transport \cite{vatratesHUN}. The calculations presented here are based on an unweighted version of the Hungarian production network.
	
We further compare the topology of the FCN with a human-to-human communication network (HCN). To this end we use a dataset provided by the same phone provider. It contains CDRs of calls between individual mobile phones which are anonymized with a new key every 24 hours. For this reason we can only analyze the HCN of one day. We choose September 17, 2020, a Thursday during the observation period outside of the holiday season and before the winter lock-downs. On that day we find 144,516 active devices and 154,557 calls.
	
We use input-output tables containing information on how many intermediate goods or services were used for the overall production of a certain good in a national economy in a given year. We use the input-output table of 2017, it is the latest available of the country studied. It contains 64 sectors in the CPA classification (\emph{Classification of Products by Activity} \cite{cpaclassification2006}), which is harmonized with NACE 2008 on level 2.
	
{\bf Systemic risk.}
We define the relative output level of firm $i$ at time t as $h_i(t) = \frac{x_i(t)}{x_i(0)}$, where $x(t)$ is firm $i$'s output at time $t$.
Let an initial firm $i$ default by setting $h_i(0)=0$. Subsequently the shock from firm $i$'s default propagates downstream along the out-links by updating all other firms' output according to their production function
\begin{equation}\label{eq:matmethods_downstream}
	x_l^{\text{d}}(t+1)  =    f_l\Big(\sum_{j=1}^{n} W_{jl} h_j^\text{d}(t) \delta_{p_j,1},  \dots,  \sum_{j=1}^{n} W_{jl}h_j^\text{d}(t) \delta_{p_j,m}\Big) \quad , 
\end{equation}
and upstream along the in-links by updating 
\begin{eqnarray}\label{eq:matmethods_upstream}
	x_l^{\text{u}}(t+1) & = &    \sum_{j=1}^{n} W_{lj}h_j^\text{u}(t) \quad. 
\end{eqnarray} 
At time $T$, the algorithm has converged and we define the vector $h_j(T)=\min(h_j^d(T),h_j^u(T))$ to calculate the economic systemic risk index as
	\begin{equation} \label{eq:matmethods_esri}
	{\rm ESRI}_{i} = \sum_{j=1}^{n} \frac{s_j }{\sum_{l=1}^n s_l }\big(1-h_j(T) \big) \quad ,
	\end{equation}
	where $s_i$ denotes the size of firm $i$. 
	For more details on the algorithm and the definition of the production function, see SI Text 4.

\section*{Acknowledgements}
We thank P. Klimek for inspiring discussions in the initial phase of the work. We thank the anonymous mobile phone operator for providing the communication data, the Hungarian National Bank for providing the Hungarian data.
The project was supported by 
Austrian Science Fund FWF under I 3073-N32,  
Austrian Science Promotion Agency FFG under 857136, and 
Hochschuljubil\"aumsstiftung of the Austrian National Bank OeNB under P17795 2018-2021

\bibliography{bibliography} 

\pagebreak
\onecolumngrid

\section*{Supplementary Information}

\FloatBarrier
\subsection*{SI Text 1: Calculating the conditional probabilities $p(s|c)$ and $p(c|s)$}

To calculate conditional probabilities describing the overlap of the communication and supply layer, we use data from a survey conducted in the country of the FCN that was conducted between April 8 and 20, 2020. The survey asked questions on the general business results of the past half year, the outlook for the near future and how the firms expected to be affected by the COVID-19 crisis in the near future.
Additionally, the survey contained a part asking for the ten most critical suppliers and ten most critical customers. From this survey we construct a supply network, that we can compare with the FCN. The survey was sent out to 102,386 companies; more than 5,955 firms replied to the supply network part, declaring more than 17,393 customer-supplier relations. To keep the privacy of the companies, the data is co-anonymized. This means that the metadata is made available to all parties prior to the data collection process and, subsequently, only anonymized data is made available to the researchers.

We quantify the overlap to find a link $x_{ij}$, $x \in \{c,s\}$, in one layer when a link $y_{ij}$, $y \in \{s,c\}$, in the other layer is present as the conditional probability $p(x|y)$,
\begin{equation}
p(x|y) = \frac{p(x \cap y)}{p(y)} \mathrm{.}
\label{eq:cond_prob}
\end{equation}
When comparing the results of the supply chain survey and the firm communication network there is an additional distinction between firms reporting in the survey and firms reported in the survey. 
Let's denote the buyer-supplier network as the set $\mathcal{S}$ containing all links $s_{ij}$ from reporting firm $i$ to reported firm $j$. This means 
\begin{equation}
i \in \mathcal{R},\qquad j \in \mathcal{M} \qquad \mathrm{ and }\qquad \mathcal{S} = \mathcal{R} \cup \mathcal{M} \quad\mathrm{,}
\end{equation}
where $\mathcal{R}$ is the set of reporting nodes, $\mathcal{M}$ the set of reported (mentioned) nodes and $\mathcal{S}$ the set of all nodes in the network. Please note that the sets $\mathcal{R}$ and $\mathcal{M}$ are not mutually exclusive; a reporting node can also be mentioned by another firm and, hence, be part of both sets.

Let's denote the communication network as the set $C$ of links $c_{ij}$ from firm i to firm j, where $i,j \in \mathcal{C}$, $\mathcal{C}$ being the set of firms in the communication network.

We are interested in the conditional probability of finding a buyer-supplier relationship where there is a communication link. Formally $p(s_{ij}|c_{ij})$. To estimate this, we need to perform a fair comparison and stratify for the fact that only a subset of all buyer-supplier relationships is known. 
\begin{figure}[ht]
	\centering
	\includegraphics[width=0.33\textwidth]{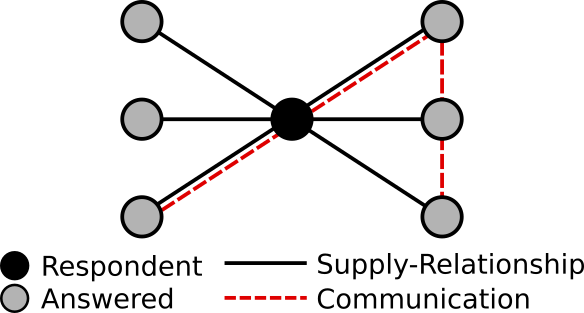}
	\caption{Schematic visualization of how to calculate conditional probabilities using a supply chain survey. By construction only supply links (solid black line) between the reporting (black) and reported (grey) nodes are observable. This means the communication links (broken red lines) need to be limited to links between reporting and reported nodes.}
	\label{fig:SI_validation_schematic}
\end{figure}
Figure SI Fig. \ref{fig:SI_validation_schematic} illustrates the problem. The central (black) node has reported in the survey and mentioned 6 other nodes (grey). We can only observe links between black and grey nodes, not between two grey nodes, so to perform a fair comparison communication links between two grey nodes need to be excluded from the analysis.
We define $\tilde{\mathcal{C}}$ in analogy to the buyer-supplier network as the set of communication links from reporting nodes to mentioned nodes, thereby excluding links between i.e. mentioned nodes
\begin{equation}
\tilde{\mathcal{C}} = \{c_{ij} | i \in \mathcal{R} \land j \in \mathcal{M}\} \mathrm{.}
\end{equation}
Now, following Eq. \eqref{eq:cond_prob} we can write
\begin{equation}
p(s_{ij}|c_{ij}) = \frac{| \tilde{\mathcal{C}} \cap \mathcal{S} |}{|\tilde{\mathcal{C}}|} \mathrm{.}
\end{equation}

To establish indicators for the error of the survey we perform a bootstrap-like simulation. Our simulations corrects for the fact that the distribution of call durations $d_{ij}$ for the subsample in the survey is not representative of the full network.
We start by sampling a synthetic supply network based on the empirical communication network using $p(s|c)$ as found in the survey. We use averages of $p(s|c)$ on bins of $d_{ij}$. Then we draw subsamples of the sample size of the survey ($N\propto 200$). After repeating the process 1500 times, we calculate mean and quartiles and report them in main text Fig. \ref{fig:layer_similarity}b.

\begin{figure}[ht]
	\centering
	\includegraphics[width=0.33\linewidth]{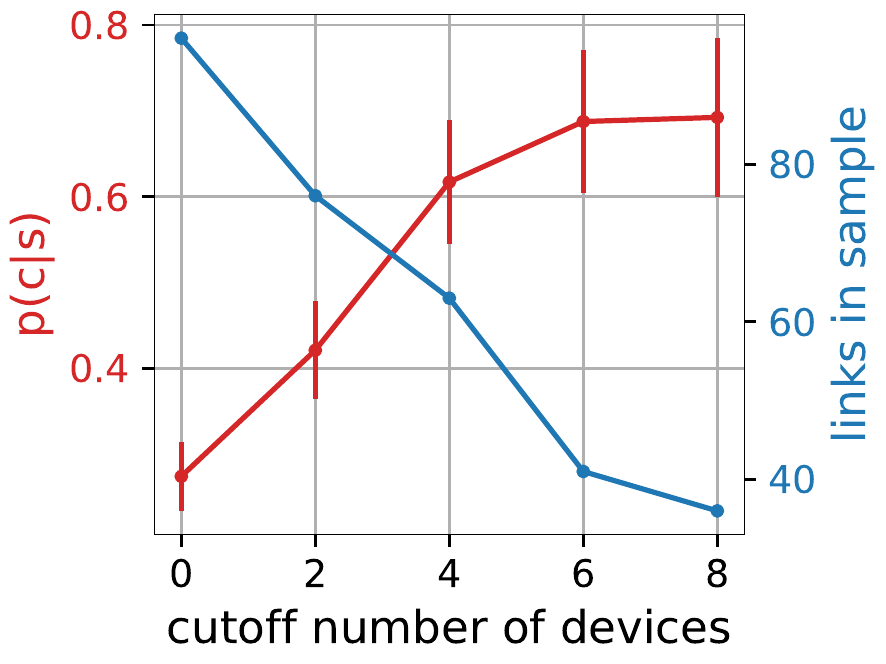}
	\caption{The conditional probability $p(c|s)$ to find a communication link if there is a supply link. We show $p(c|s)$ as a function of the firm size thresholds, which are chosen as proxies for the supply link strength. The blue line shows the number of supply links in the sample used to calculate $p(c|s)$.}
	\label{fig:SI_psc}
\end{figure}

For $p(c|s)$ we lack the true distribution of supply link strengths, so we perform a classic bootstrap. For a given conditional probability bin we draw samples of the same size with replacement. We repeat the process 1500 times, calculate mean and quartiles and report them in SI Fig. \ref{fig:SI_psc}.

\FloatBarrier
\subsection*{SI Text 2: Finding the optimal threshold using Kullback-Leibler divergence}

The overlap probabilities $p(s|c)$ and $p(c|s)$ increase when thresholded for call duration and/or firm sizes. 
Of course on the one hand, when thresholding, the information contained in low-intensity contacts (low call duration, small trade volumes) is lost, while on the other hand, link correlations between the communication and supply layers increase. To balance these two effects, we choose the threshold combination where the topology of the thresholded communication network is most similar to a real production network.

\begin{figure}[ht]
	\centering
	\includegraphics[width=0.66\textwidth]{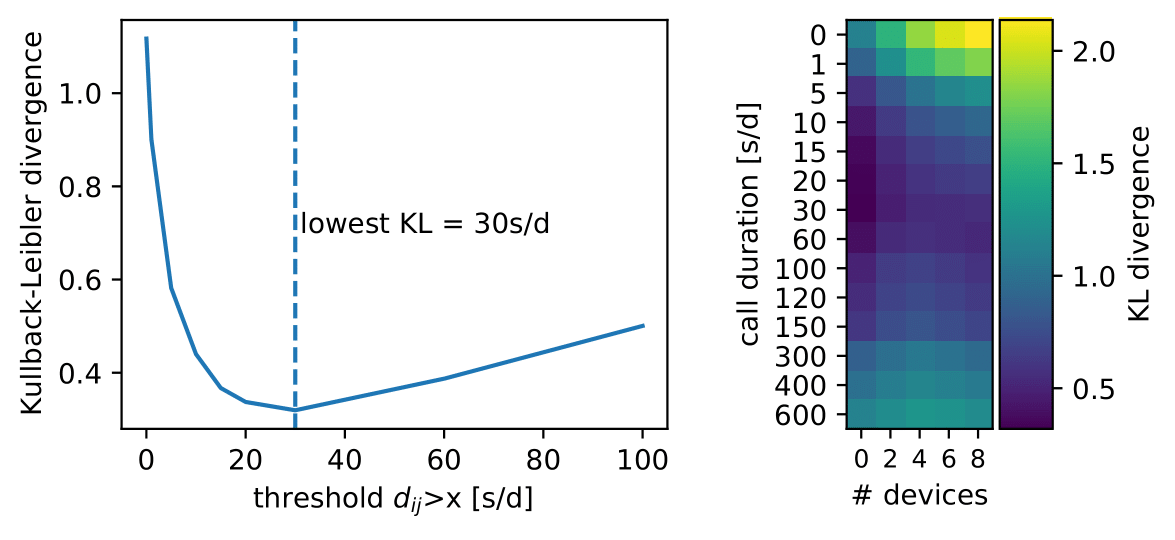}
	\caption{Finding the optimal threshold combination by minimizing the Kullback-Leibler divergence. (a) Kullback-Leibler divergence between $p(k)$ of the HSN and the FCN for different thresholds of the average call duration $d_{ij}$. The lowest value for KL is at $d_{ij}=30\mathrm{s/d}$. (b) Heatmap showing KL between the HSN and the FCN for different thresholds of $d_{ij}$ and the number of devices $N_i$. KL is lowest for $d_{ij}>30\mathrm{s/d}$ and $N_i>0$.}
	\label{fig:SI_KLdivergence}
\end{figure}

To determine when the topology of the thresholded FCN most resembles a real production network, we calculate the Kullback-Leibler divergence between the thresholded FCN and the HSN,
\begin{equation}
KL = \sum_k p(k^{FCN}) \mathrm{log} \left( \frac{p(k^{FCN})}{p(k^{PNW})} \right) \mathrm{.}
\end{equation}
We systematically try threshold combinations for the average call duration per week $d_{ij}$ and the number of devices per firm $N_i$, $(d_{ij},N_i)$. As shown in SI Fig. \ref{fig:SI_KLdivergence}a and \ref{fig:SI_KLdivergence}b, the minimal Kullback-Leibler divergence is found for $(30\mathrm{s/d},0)$. 

\FloatBarrier
\subsection*{SI Text 3 Comparing the network topologies of the FCN, HSN and HCN}

Here we compare the network topologies of the FCN, HSN and HCN we discuss the behavior of $p(k)$, $k_{nn}(k)$ and $c(k)$ in greater detail. 
We begin by characterizing the topology of the firm communication network (FCN). For reference we compare the results to a more directly observed supply network and to a human communication network and show that the topology of the FCN is similar to the known supply network and dissimilar to the social network.
%
Here we describe the network thresholded to only links between firms that have an average interaction duration of more than 30 seconds per day.
For the comparison with a real supply network, we compare with the national supply network of Hungary that is obtained through VAT data for 2017 \cite{borsos2020unfolding, diem2021quantifying} (henceforth HSN, short for ``Hungarian supply network''). The network shows a link if the tax content of the goods exchanged between two firms exceeds 1,000,000 Forint (approx. 3,000 Euro) and if the link occurs in at least two distinct quarters. For details on the supply network data, see Materials and Methods and \cite{borsos2020unfolding}.
To investigate the difference between the inter-firm communication network and a social communication network (HCN) between humans we use a dataset on calls of individual devices by the same mobile phone provider. The data is for one day during the studied period because the IDs of individual devices are re-anonymized daily, preventing us from studying the communication network for longer than 24h. For details on the HCN we refer to Materials and Methods. 

In Fig. \ref{fig:layer_similarity}a we show the degree distribution $p(k)$ of the three networks. In complex networks the degree distribution is often skewed to the right with a fat tail, corresponding to the fact that a few hubs are interacting with many nodes, while the majority of nodes interacts with only a few nodes.
The FCN has an average degree of $\langle k^{FCN} \rangle = 4.79$. Its degree distribution has a maximum at $k^{FCN}=2$ and has a fat tail that can asymptotically be described by a power-law 
\begin{equation}
p(k) \propto k^{-\alpha} \mathrm{,}
\label{eq:degreepowerlaw}
\end{equation}
which we fit with a maximum likelihood estimator using Python's \emph{powerlaw} package \cite{alstott2014powerlaw}.
We find $\alpha_k^{FCN} = 2.18(12)$ for $k^{FCN}>30$.
The PNW does not show the increase for small $k$ but also exhibits a fat tail that can be well fitted by the power law from Eq. \ref{eq:degreepowerlaw} with $\alpha_k^{HSN} = 2.40(3)$ for values of $k^{HSN}>30$. The average degree is  $\langle k^{HSN} \rangle = 2.1$. For the Japanese production network an average degree of $\langle k^{JPN} \rangle = 8.0$ and degree exponents of $\alpha_k^{JPN,in} = 2.35$ and $\alpha_k^{JPN,out} = 2.26$ have been reported \cite{fujiwara2010large}.

For the HCN we find an average degree of  $\langle k^{HCN} \rangle = 4.75$. There the decrease of $p(k)$ for high values is stronger, with a degree exponent of $\alpha_k^{HCN} = 4.89(26)$. Our results are between the exponents found in two studies in the literature. For a statistically validated communication network in Shanghai, Li et al. fit exponentially truncated power laws and report $\alpha_k^{in} = 2.76$ and $\alpha_k^{out} = 2.90$ for the in- and out-degree, respectively \cite{li2014comparative}. Onnela et al. report for a mobile phone communication network in Spain, where they find a tail exponent of $\alpha_k = 8.4$ \cite{onnela2007analysis}. The difference can be perhaps explained by a change in the way mobile phones are used, as, Li et al. argue, a claim which is also supported by a significantly lower degree exponent of $\alpha_k = 2.1$ for landlines, as Onnella et al. report.

The mixing patterns of a network have a strong influence on its structure and function. If high-degree nodes tend to interact with other high-degree nodes, the network is called assortative, if high-degree nodes are more likely to interact with low-degree nodes, the network is called disassortative. For supply networks disassortative mixing has been reported \cite{fujiwara2010large}; for social networks we expect assortative mixing \cite{onnela2007analysis,li2014comparative}.
To investigate the degree-degree correlations we plot the average nearest neighbor degree 
\begin{equation}
\langle k_{nn} \rangle = (1/k_i) \sum_{j\in \mathcal{N}_i} k_j \textrm{,}
\end{equation} 
where $\mathcal{N}_i$ is the set of neighbor nodes of $i$.
In Fig. \ref{fig:layer_similarity}b we plot $\langle k_{nn} \rangle$ as a function of $k$ for the FCN (orange), PNW (blue) and HCN (green). For the FCN $\langle k_{nn} \rangle(k)$ shows an increase for small values below $k<10$ and then shows decreasing trend for larger $k$. For very small values the network is assortative, for intermediate and large $k$ the firm communication network is disassortative.
For the PNW we find $\langle k_{nn} \rangle$ to be relatively flat for small $k<10$ and then decrease for large $k$, thereby showing that the PNW is disassortative for $k>10$. This result agrees well with previous studies on the Japanese supply network, which was also shown to be disassortative.
The HCN increases to values around $k \approx 30$ and then decreases quickly, suggesting the existence of two regimes; a low-degree regime, where assortative mixing patterns dominate, and a high-degree regime, where disassortative mixing is predominant. For the mobile phone communication network of Shanghai very similar results were found. There the authors associate the assortative mixing for nodes with `reasonable degree' to the social network of calls and the disassortative mixing of high-degree nodes with hotlines or robots \cite{li2014comparative}. This suggests, that for `human' callers, the network is assortative.

The local cohesiveness around one node is typically measured by the local clustering coefficient $c_i$. It is defined as the number of closed triangles of node i with its neighbors, $t_i$, divided by the number of possible triangles \begin{equation}
c_i = \frac{2 t_i}{k_i(k_i-1)} \textrm{.}
\end{equation} 
The average local clustering coefficients $\langle c \rangle$, along with their expected value for a random network $p$ is shown in Tab. \ref{tab:nwdescription}. All clustering coefficients are comparably small, but still large compared to random networks.
Figure \ref{fig:layer_similarity}c shows average $c_i$ as a function of degree $c(k)$. For the FCN (orange) and the HSN (blue) $c(k)$ shows a very similar decay.
The HCN (green) does not show such a decay, but a decrease for values below $k\approx20$ and an increase for values between $k\approx20$ and $k\approx40$. This weak, peaked dependence of $c$ on $k$ for a mobile phone network was also reported in \cite{li2014comparative}.

\begin{table}
	\centering
	\caption{Network characteristics for four networks in our study. We show the average degree $\langle k \rangle$, the average clustering coefficient $\langle c \rangle$ the linking probability for a random network of the same density $p$ and the average nearest neighbor degree $\langle k_{nn} \rangle$ for the unthresholded FCN, for the RSN, the HCN and the HSN.}
	\begin{tabular}{l|c|c|c|c}
		Network & $\langle k \rangle$ & $\langle c \rangle$ & $p$ & $\langle k_{nn} \rangle$ \\
		\hline
		FCN  (no threshold) & 23.4 &  0.24 & 0.0024 & 516\\
		
		RSN  ($d_{ij} > 30\mathrm{s/d}$) & 4.79 &  0.09 & 0.0006 & 77\\
		
		HCN  & 2.1 & 0.09 & 0.000015 & 3.7 \\
		
		SNW  & 4.75 & 0.06 & 0.000052 & 157 \\
		
	\end{tabular}
	\label{tab:nwdescription}
\end{table}

\FloatBarrier
\subsection*{SI Text 4: Calculating Economic Systemic Risk}
Here we describe how the ESRI is calculated. We keep our notation closely to the one of \cite{diem2021quantifying}.

Given a supply network $W$, where $W_{ij}$ describes the value of products, of type $p_i$, delivered from firm $i$ to firm $j$. The vector $p$ with element $p_i\in \{1,2, \dots, m\}$ indicates the product type produced by firm $i$. We identify the product $p_i$ with a firm's industry affiliation. The amount of input $k$ firm $j$ uses is $	\Pi_{jk}=\sum_{i=1}^{n} W_{ij}\delta_{p_i,k}$.   A firm's production function is an increasing function $f_i(\Pi_{i1},\dots,\Pi_{im})$ that describes how much firm $i$ can produce with a given set of inputs $\Pi_{i1},\dots,\Pi_{im}$.  Conversely it also allows to assess how much production drops if the amount $\Pi_{ik}$ of input type $k$ is reduced. 

The generalized Leontief production function is a generalization of the regular Leontief production function --- with functional form $x_i = \min\Big( \frac{1}{\alpha_{i1}}\Pi_{i1}, \frac{1}{{\alpha_{i2}}}\Pi_{i2}\Big) \quad $ --- and a linear production function --- with functional form $x_i = \frac{1}{\alpha_{i1}} \Pi_{i1} + \frac{1}{\alpha_{i2} }\Pi_{i2} \quad $ --- and is defined as
\begin{eqnarray} \label{eq:def_modified_leontief_pf}
	x_i & = & \min \Bigg[
	\min_{k \in \mathcal{I}_i^\text{es}} \Bigg( \frac{1}{\alpha_{ik}}\sum_{j=1}^{n}  W_{ji} \delta_{p_{j},k}\Bigg), \:\beta_{i} + \frac{1}{\alpha_i} \sum_{k \in  \mathcal{I}_i^\text{ne}}  \sum_{j=1}^{n}  W_{ji} \delta_{p_{j},k} \Bigg]
\end{eqnarray}
where the set, $\mathcal{I}_i^\text{es}$, denotes all input types $k$, that are deemed essential for production of firm $i$ and thus entering the production in a Leontief way,  the set $\mathcal{I}_i^\text{ne}$ denotes all input types $k$ that enter the production of firm $i$ in a linear way. 
The parameter $	\alpha_{ik} = \frac{ \sum_{j=1}^{n}  W_{ji} \delta_{p_{j},k} }{ \sum_{l=1}^{n}  W_{il} }$ is the fraction of firm $i$'s output that it spends on the input type $k$, the parameter $	\alpha_i = \frac{\sum_{j=1}^{n} W_{ji}}{\sum_{l=1}^{n} W_{il}} $ is the overall fraction of output that is spend on all inputs and
$\beta_i$ is another parameter inferred from the supply network and defined as the attainable production level if only essential inputs $k \in \mathcal{I}_i^\text{es}$ are available, i.e.,
\begin{equation}
	\beta_i = \Bigg(\sum_{l=1}^{n} W_{il} \Bigg) \frac{ \sum_{k \in  \mathcal{I}_i^\text{es}}  \sum_{j=1}^{n}  W_{ji} \delta_{p_{j},k} }{  \sum_{j=1}^{n}  W_{ji}  } \quad. 
\end{equation} \\

\noindent

We list the necessary equations to compute the ESRI. 
First, the following objects have to be defined. The downstream impact matrix $\Lambda^\text{d}$ defined by
\begin{equation}  \label{eq:algorithm_lambda_d}
	\Lambda_{ij}^\text{d} = \begin{cases}
		\Lambda^{d1}_{ji} \qquad \text{if} \; p_j \in \mathcal{I}_i^\text{es} \\
		\Lambda^{d2}_{ji} \qquad \text{if} \; p_j \in \mathcal{I}_i^\text{ne}
	\end{cases} \, ,
\end{equation}
and the elements of $\Lambda^{d1}$  and $\Lambda^{d2}$  are defined as  
\begin{align} \label{eq:algorithm_lambda_d1_d2} 
	\Lambda^{d1}_{ji} =  \begin{cases}  
		\frac{W_{ji}}{\sum_{\iota=1}W_{\iota i}  \delta_{p_{\iota},p_{j}} } \quad \text{if} \; W_{ij} > 0 \qquad , \\
		\qquad \quad 0 \quad \qquad \; \text{else} \quad\qquad\qquad,
	\end{cases} \\
	\Lambda^{d2}_{ji} =  \begin{cases}  
		\frac{W_{ji}}{\sum_{l=1}W_{li}   } \qquad \quad \text{if} \; W_{ij} > 0 \qquad , \\
		\qquad \quad 0 \quad \qquad \; \text{else} \quad\qquad\qquad.
	\end{cases}
\end{align} 
Similarly, the upstream impact matrix is defined by
\begin{eqnarray} \label{eq:lambda_u}
	\Lambda^\text{u}_{ji} = 
	\begin{cases} \frac{W_{ij}}{ \sum_{l=1}^n W_{il} } \qquad \text{if} \quad  W_{ij}>0 \quad,  \\
		\qquad  0 \qquad \quad \text{else} \qquad \qquad \; \;  , \end{cases}
\end{eqnarray}

Second, to calculate the ESRI of firm $i$ the initial exogenous shock parameter is chosen to be $\psi_i=0$ and $\psi_j=1$ for the other firms $j\neq i$. 

Third, the following equations are iteratively computed to update the downstream and upstream impeded production levels of firms at time point $t$. 
\begin{enumerate}
    \item Compute the dynamic intraindustry market share for each firm $j$
    \begin{equation} \label{eq:algorithm_sigma}
	\sigma_j(t) =  \min \Big(\frac{s^\text{out}_j(0)}{\sum_{l=1}^{n} s^\text{out}_l(0)h_l^d(t) \delta_{p_l,p_j}},1 \Big) \quad. 
\end{equation}
\item Compute, for essential inputs $k\in \mathcal{I}_i^\text{es}$ of firm $i$ the fraction available as
\begin{equation} \label{eq:algorithm_rel_input_essential}
	\tilde{\Pi}_{ik}(t) = 1 - \sum_{j=1}^n \sigma_j(t) \Lambda^{d}_{ji} \big(1-h_j^\text{d}(t) \big) \: \delta_{p_j,k} \quad .
\end{equation}
\item Compute, for non-essential inputs, $k\in \mathcal{I}_i^\text{ne}$, of firm $i$ the fraction available as
\begin{equation} \label{eq:algorithm_rel_input_non_essential}
	\tilde{\Pi}_{ik'}(t) = 1 - \sum_{k\in \mathcal{I}_i^\text{ne}}  \sum_{j=1}^n \sigma_j(t) \Lambda^{d}_{ji} \big(1-h_j^\text{d}(t) \big) \: \delta_{p_j,k} \quad .
\end{equation}
\item Update for each firm $i$ the relative production level reduced by downstream shocks
\begin{equation} \label{eq:algorithm_downstream_update}
	h_i^{\text{d}}(t+1)   =   \min\Big[
	\min_{k \in \mathcal{I}_i^\text{es}} \Big(\tilde{\Pi}_{ik}(t)   \Big), \;
	\tilde{\Pi}_{ik'}(t)
	, \psi_i \Big] \quad ,
\end{equation}

\item  Update for each firm $i$ the relative production level reduced by upstream shocks
\begin{equation} \label{eq:upstream_update}
	h^u_{i}(t+1) = \min \Big[ \sum_{j=1}^n \Lambda^\text{u}_{ji}h^\text{u}_j(t), \psi_i  \Big] \quad.  
\end{equation}
\end{enumerate}
 
The iteration continues until the algorithm reaches a stable state at time
\begin{equation}
	T \equiv \min_t\{t \in \mathbb{N}|\max\big(  h^d(t)-h^d(t+1) ,   h^u(t)-h^u(t+1)   \big) \leq \epsilon \} + 1 \quad ,
\end{equation}
with $\epsilon=10^{-2}$ as convergence threshold. 

Then the ESRI$_i$ of firm $i$ is computed as
\begin{equation} \label{eq:matmethods_esri_SI}
	{\rm ESRI}_{i} = \sum_{j=1}^{n} \frac{s_j }{\sum_{l=1}^n s_l }\big(1-h_j(T) \big) \quad .
\end{equation}
The quantity can be interpreted as the fraction of production in the network that is (temporarily) impeded if firm $i$ fails (temporarily). 

For details of the derivation see \cite{diem2021quantifying} Appendix G.

Note that the calculation is computationally intensive and scales badly, because with growing network the number of ESRI$_i$ to calculate grows linearly and the convergence times grows by 2 times matrix multiplication costs. For this reason we only show the 1000 most risky firms in Figure \ref{fig:ESRIspreadMain} (d).

\FloatBarrier
\subsection*{SI Text 5: Extended analysis of the ESRI profiles}
\begin{figure*}[ht]
	\centering
	\includegraphics[width=0.4\textwidth]{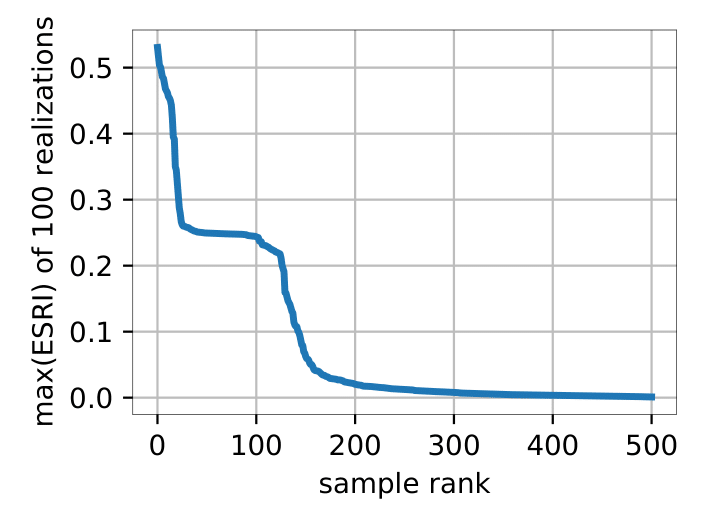}
	\caption{Maximal worst-case economic systemic risk. We plot the maximal ESRI per node found after simulating 100 configurations of the RSNW ordered from highest to lowest. The shown values are the maximal damage a node does in 100 scenarios, they do not occur all together in one configuration.}
	\label{fig:SI_ESRImax}
\end{figure*}

We are not only interested in the median damage a company can do, but also in the scenario where the damage is largest. In SI Fig. \ref{fig:SI_ESRImax} we plot the maximal ESRI per node of 100 reconstructed supply networks. Reconstructing 100 supply networks and calculating ESRI yields a distribution for every node. Supplementary Figure \ref{fig:SI_ESRImax} shows the maximum of each distribution. Note that the maxima are not all from the same configuration. The maximal damage is $max(ESRI)= 0.53$ and the high systemic risk core consists of around 100 firms, with the majority of nodes having an ESRI around $ESRI\approx 0.25$.

\begin{table}
	\centering
	\caption{NACE lvl. 2 classification of the firms in the high systemic risk plateau.}
	\begin{tabular}{l|l|r|l}
		code & sector name &  frequency & percent \\
		\hline
		C28  &                                                                                    Manufacture of machinery and equipment n.e.c. &         10 &   15.4\% \\
		C17  &                                                                                          Manufacture of paper and paper products &          6 &    9.2\% \\
		D35  &                                                                              Electricity, gas, steam and air conditioning supply &          5 &    7.7\% \\
		C16  &  Manufacture of wood and of products of wood and cork,... &          5 &    7.7\% \\
		C23  &                                                                               Manufacture of other non-metallic mineral products &          4 &    6.2\% \\
		C10  &                                                                                                     Manufacture of food products &          4 &    6.2\% \\
		C22  &                                                                                       Manufacture of rubber and plastic products &          4 &    6.2\% \\
		C25  &                                                         Manufacture of fabricated metal products, except machinery and equipment &          4 &    6.2\% \\
		K64  &                                                               Financial service activities, except insurance and pension funding &          4 &    6.2\% \\
		C11  &                                                                                                         Manufacture of beverages &          2 &    3.1\% \\
		C24  &                                                                                                      Manufacture of basic metals &          2 &    3.1\% \\
		C20  &                                                                                   Manufacture of chemicals and chemical products &          2 &    3.1\% \\
		C29  &                                                                        Manufacture of motor vehicles, trailers and semi-trailers &          2 &    3.1\% \\
		C32  &                                                                                                              Other manufacturing &          2 &    3.1\% \\
		F43  &                                                                                              Specialised construction activities &          2 &    3.1\% \\
		A01  &                                                               Crop and animal production, hunting and related service activities &          1 &    1.5\% \\
		G47  &                                                                           Retail trade, except of motor vehicles and motorcycles &          1 &    1.5\% \\
		E38  &                                                          Waste collection, treatment and disposal activities; materials recovery &          1 &    1.5\% \\
		C33  &                                                                               Repair and installation of machinery and equipment &          1 &    1.5\% \\
		C31  &                                                                                                         Manufacture of furniture &          1 &    1.5\% \\
		F42  &                                                                                                                Civil engineering &          1 &    1.5\% \\
		C21  &                                                     Manufacture of basic pharmaceutical products and pharmaceutical preparations &          1 &    1.5\% \\
	\end{tabular}
	\label{tab:sector_composition_lvl2}
\end{table}

\FloatBarrier
\subsection*{SI Text 6: Limitations for systemic risk calculation}
Our study is subject to several limitations, in particular (i) the error due to faulty direction/weight estimation, (ii) the limited market coverage of the phone provider (resulting in limited agreement even if $p(s|c) = p(c|s) = 1$) and (iii) the imperfect overlap of the two networks limiting the possible accuracy. In the following, we address all of these shortcomings one by one and discuss the size of the introduced biases and errors. We end with a simulation to estimate the error introduced by all shortcomings combined.

\begin{figure}[ht]
	\centering
	\includegraphics[width=0.99\linewidth]{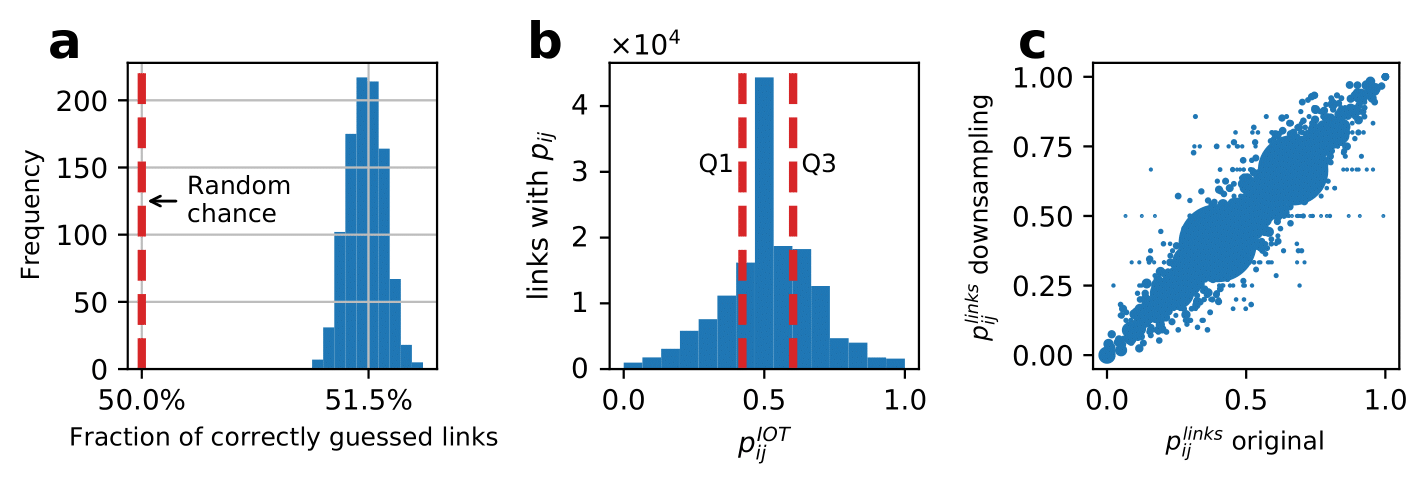}
	\caption{Accuracy of the link direction estimation. (a) We aggregate the weights of the HSN to a NACE lvl. 2 input-output table. From this we compute $p_{ij}$ and estimate the link direction for every link. We calculate the fraction of correctly guessed link directions and show the result of 1000 iterations in the blue histogram. The average fraction of correctly guessed links is 51.5\%. Using no additional information from input-output tables, i.e. assuming $p_{ij} = 0.5$ would result in an average fraction of correctly guessed links of 50\% (red vertical line).
	(b) Number of links associated with a given value of the direction probability $p_{ij}^{IOT}$. The distribution is centered around 50\%, indicating that for the majority of the links there is no clear flow direction. The red vertical lines indicate the first and third quartile at 42\% and 60\%, respectively. The peak at 50\% is because many trade links are inside one sector, resulting in $p_{ii}=0.5$.
 	(c) Fraction of links $p_{ij}$ from i to j in the original network and after downsampling the links from the aggregated network for one reconstruction. Although, as shown in panel (a), the individual directions are not captured very well, on the aggregate level the fraction of links pointing from i to j after reconstruction correlates strongly with the values from the original network. After 1000 iterations we find an average Pearson correlation of $\langle r \rangle =0.91(1)$.}
	\label{fig:pij_frac_corr}
\end{figure}

We study the error introduced by sampling the link directions by simulating the reconstruction process on a real supply network topology and then validating it by comparison with the true network. We use the Hungarian supply network and start by aggregating the trade volumes to an input-output table. Then we remove the direction information from the network and sample new link directions according to the probabilities obtained from the input-output table and Eq. \eqref{eq:pij_iot}.   
We compare the true and simulated link directions and calculate which fraction was guessed correctly. Figure \ref{fig:pij_frac_corr}a shows results of repeating this experiment 1000 times. The mean overlap is 51.5\% with a standard deviation of 0.1\%.
This result, although significantly better than what would be expected from random chance (red line in Fig. \ref{fig:pij_frac_corr}b), is surprisingly low. It can be explained by investigating the probabilities associated with the links in Hungary. If most links were between sectors with a direction as polarized as the relationship between agriculture and the food industry, more links would be guessed correctly. However, as shown in SI Fig. \ref{fig:pij_frac_corr}b the majority of links has probabilities between 42\% and 60\% (the lower and upper quartile, shown as red lines).   
Nevertheless, even though many direct links are guessed incorrectly, on the sector level the proportion of links from sector i to sector j $p_{ij} = L_{ij}/(L_{ij}+L_{ji})$ are captured well. We find an average Pearson correlation of the true, empirical sector wise link directions $p_{ij}^{emp}$ and the simulated sector wise link directions $p_{ij}^{sim}$ of $\langle r(p_{ij}^{emp},p_{ij}^{sim}) \rangle = 0.91(1)$, see also SI Fig. \ref{fig:pij_frac_corr}c.

\begin{figure*}[ht]
	\centering
	\includegraphics[width=0.95\textwidth]{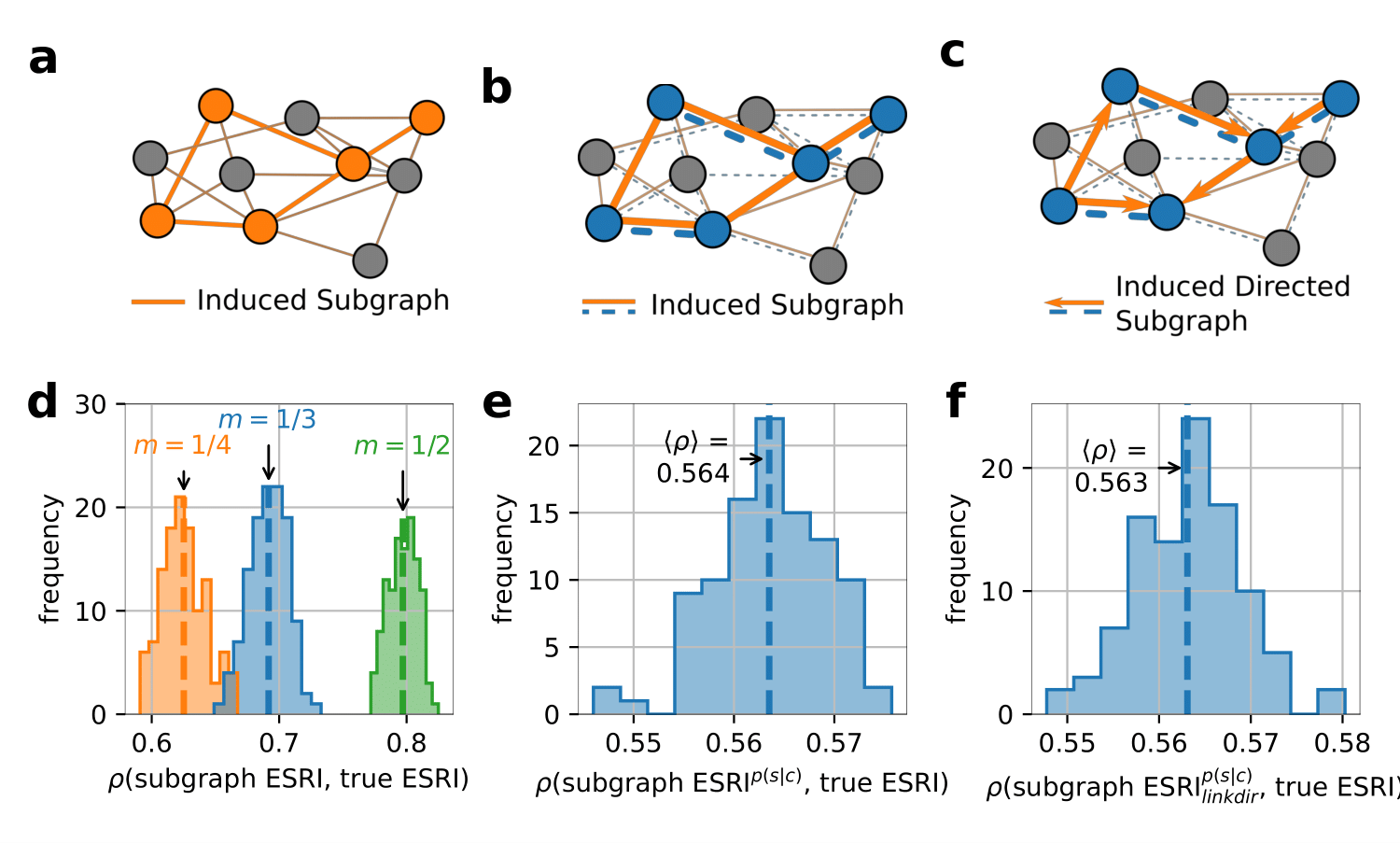}
	\caption{Limitations for systemic risk calculations. Perform simulations on the HSN to estimate the size of the error introduced by considering (a) the induced subgraph of calls between customers of a mobile phone company with market share $m$, (b) the induced subgraph on a multilayer network with imperfect overlap and (c) the induced subgraph on a multilayer network with imperfect overlap where the link directions need to be estimated. (d) The correlation coefficient of the ESRI calculated on a full network and a subgraph of size $m=1/3$ (blue) and $m=1/2$ (green). The histograms show the results of 120 iterations and the vertical links highlight the median values $\langle \rho \rangle_{m=1/4}=0.63(2)$,
		$\langle \rho \rangle_{m=1/3}=0.69(1)$ and 
		$\langle \rho \rangle_{m=1/2}=0.80(1)$.
		(e) On top of the induced subgraph we limit the overlap of the communication and supply layer by generating a synthetic communication layer using $p(s|c)=0.21$. The correlation coefficients of 100 simulations with $m=1/3$ are shown in the blue histogram, the mean correlation coefficient is $\langle \rho \rangle = 0.564(5)$ (red vertical line).
		(f) As an additional step we add the reconstruction of the link directions as described in the main text. The correlation coefficients of 100 simulations are shown in the blue histogram, the mean correlation coefficient is $\langle \rho \rangle = 0.563(6)$ (red vertical line).
		Arguably the largest contribution to the overall error is caused by the limited market share, followed by limited overlap $p(c|s)<1$. The error introduced due to the link direction estimation is only marginal.
	}
	\label{fig:bias_est}
\end{figure*}

As in most countries, there is more than one mobile phone provider in the country where the mobile phone data is from. This results in a market share $m$ less than one. As is schematically shown in Fig. \ref{fig:bias_est}a, this leads to a large fraction of links that are not accounted for, since we only consider calls between companies who are customers of the mobile phone provider.
The graph containing only the links between a set of nodes in a network is called the \emph{induced subgraph} of the respective set of nodes.
To quantify the error introduced by limited coverage we use the real supply network of Hungary and compare the systemic risk as calculated on the full network with the systemic risk calculated on an induced subgraph. We investigate the effect of a market coverage of $m=1/4$, $m=1/3$ and $m=1/2$ using the following steps.
\begin{enumerate}
	\item Calculate ESRI for the full Hungarian network.
	\item Draw a sample of nodes according to market share $m$, calculate the induced subgraph.
	\item Calculate ESRI on induced subgraph.
	\item Correlate ESRI of induced subgraph with ``true" ESRI of these nodes.
	\item Repeat from 2. and calculate the histogram of correlation coefficients.
\end{enumerate}    
Figure \ref{fig:bias_est}b shows the results of 120 iterations, the average Spearman correlation coefficient $\rho(ESRI_{full}, ESRI_{subgraph})$ for 
$m=1/4$ (orange) is $\langle \rho \rangle_{m=1/4}=0.63(2)$,
$m=1/3$ (blue) is $\langle \rho \rangle_{m=1/3}=0.69(1)$ and
$m=1/3$ (green) is $\langle \rho \rangle_{m=1/2}=0.80(1)$.

The next step is to quantify the error introduced by the imperfect link correlations, $p(s|c) < 1$ and $p(c|s) < 1$. Figure \ref{fig:bias_est}b illustrates the imperfect overlaps of the communication (broken blue lines) and the supply (solid orange lines) layers.
We generate a fake communication network, based on the HSN and the probabilities to find a communication link where a supply link is present $p(c|s)$ and where no supply link is present $p(c|\lnot s)$. It is not possible to measure $p(c|\lnot s)$ directly. Nevertheless we can compute it from known quantities
\begin{equation}  
p(c|\lnot s) = \frac{p(c) - p(c|s)p(s)}{1-p(s)} \mathrm{.}
\end{equation}
We are interested in the effect on top of the error introduced by the incomplete market coverage, therefore we sample nodes according to a market share of $m=1/3$ and calculate the induced subgraph. To isolate the effect, however, we keep the directions from the HSN and investigate their effect in the next step.
The modified algorithm works as follows:
\begin{enumerate}
	\item Calculate ESRI for the full Hungarian network.
	\item Generate "fake" mobile phone network for Hun using $p(c|s)$ and $p(c|\lnot s)$
	\item Calculate ESRI on the simulated mobile phone network.
	\item Correlate ESRI of the simulated network with the "true" ESRI.
	\item Repeat from 2. and make histogram of correlation coefficients.
\end{enumerate}

Figure \ref{fig:bias_est} show the results of 100 iterations assuming $m=1/3$ and $p(c|s) = 0.21$. The mean Pearson correlation is $\langle \rho \rangle_{p(s|c)} = 0.564(5)$, demonstrating a substantial shift of $\Delta \langle \rho \rangle = 0.13$ compared to the result not including $p(s|c)<1$.

Finally, we study the combined effect of the limitations described above, as shown in Fig. \ref{fig:bias_est}c we calculate the effects of a limited market share, imperfect link correlations and an inaccurate link direction estimation. We use the empirical network topology of Hungary, simulate a mobile phone network and then estimate the link directions. The process follows the steps below. 
\begin{enumerate}
	\item Calculate ESRI for the full Hungarian network.
	\item Generate``fake" mobile phone network for Hun using $p(c|s)$ and $p(c|\lnot s)$
	\item Draw a sample of nodes according to market share $m$, calculate induced subgraph.
	\item Reconstruct the directions using input-output tables.
	\item Calculate ESRI on induced subgraph of the simulated phone network.
	\item Correlate ESRI of induced subgraph with "true" ESRI of these nodes.
	\item Repeat from 2. and make histogram of correlation coefficients.
\end{enumerate}

Figure \ref{fig:bias_est}c shows the results for 100 iterations with $m=1/3$ and $p(c|s) = 0.21$. We find an average Pearson's correlation coefficient of $\langle \rho(ESRI_{full}, ESRI_{reconstr})\rangle = 0.563(6)$. Compared to the previous simulation the estimation of the link directions adds only a small error $\delta \langle r \rangle = 0.0004$ to the final result.	

\FloatBarrier
\subsection*{SI Text 7: Anonymization procedure}
The firm communication dataset is merged with a commercially available business intelligence database that was made available to the mobile phone provider. This database includes balance sheet information from which we proxy the firm's sizes by their total assets, and on their industry classification in the NACE 2008 system  \cite{naceclassification2006}). Because this information would potentially make it possible to identify individual firms, the merged data does not leave the premises of the phone company. All calculations were executed there and then anonymized. In this form the data was handed to us and was destroyed at the phone company after the project.

To calculate conditional probabilities describing the overlap of the communication and supply layer, we perform a large survey to obtain ground truth data, for details see SI Text 1. To keep the privacy of the firms, the data is co-anonymized. This means that metadata is exchanged between the researchers and the mobile phone provider and a shared anonymization procedure is employed. Finally, only fully anonymized data is made available to the researchers.

\end{document}